# Generation and Electric Control of Spin-Coupled Valley Current in WSe$_2$

Hongtao Yuan,[1,2] Xinqiang Wang,[3,4] Biao Lian,[1] Haijun Zhang,[1] Xianfa Fang,[3,4] Bo Shen,[3,4] Gang Xu,[1] Yong Xu,[1] Shou-Cheng Zhang,[1,2] Harold Y. Hwang,[1,2*] Yi Cui[1,2*]

[1]*Geballe Laboratory for Advanced Materials, Stanford University, Stanford, California 94305, USA*
[2]*Stanford Institute for Materials and Energy Sciences, SLAC National Accelerator Laboratory, Menlo Park, California 94025, USA*
[3]*State Key Laboratory of Artificial Microstructure and Mesoscopic Physics, School of Physics, Peking University, Beijing, 100871 China*
[4]*Collaborative Innovation Center of Quantum Matter, Beijing, China*

**The valley degree of freedom in layered transition-metal dichalcogenides (MX$_2$) provides the opportunity to extend functionalities of novel spintronics and valleytronics devices. Due to spin splitting induced by spin-orbital coupling (SOC), the non-equilibrium charge carrier imbalance between two degenerate and inequivalent valleys to realize valley/spin polarization has been successfully demonstrated theoretically and supported by optical experiments. However, the generation of a valley/spin current by the valley polarization in MX$_2$ remains elusive and a great challenge. Here, within an electric-double-layer transistor based on WSe$_2$, we demonstrated a spin-coupled valley photocurrent whose direction and magnitude depend on the degree of circular polarization of the incident radiation and can be further greatly modulated with an external electric field. Such room temperature generation and electric control of valley/spin photocurrent provides a new property of electrons in MX$_2$ systems, thereby enabling new degrees of control for quantum-confined spintronics devices.**

---

[*] Corresponding author: hyhwang@stanford.edu, yicui@stanford.edu.

Generation and manipulation of a spin current is one of the most critical steps in developing semiconductor spintronic applications.[1, 2, 3] In a two-dimensional electronic system (2DES) with spin degeneracy lifted, irradiation with circularly polarized light can result in a non-uniform distribution of photo-excited carriers in $k$-space following optical selection rules and energy/momentum conservation, finally leading to a spin current.[4, 5, 6] Referred to as the circular photogalvanic effect (CPGE),[7, 8, 9, 10] the fingerprint of such a spin photocurrent is the dependence on the helicity of the light. As schematically shown in Fig. 1a, in a Rashba 2DES,[11, 12] the absorption of circularly polarized light results in optical spin orientation by transferring the angular momentum of photons to electrons, and thus, the non-equilibrium spin polarization of electrons forms a spin current with the electron motion in the 2DES plane. Generally, the amplitude of the CPGE current can be expressed by $j_{CPGE} = \eta \gamma I \sin\theta \sin 2\varphi$,[7] where $\eta$ is the absorbance, $\gamma$ is the matrix element which is related with the spins, orbitals and symmetry of the 2DES, $I$ is the incident light intensity, $\theta$ is the incident angle (Fig. 1b), and $\varphi$ is the rotation angle of the quarter wave plate (reflecting the helicity of the incident circularly polarized radiation). This electric current has two important characteristic features: a) its direction and magnitude depend on the degree of circular polarization of the incident light, as indicated in Fig. 1b; and b) it can be controlled via modulating the $\gamma$ coefficient, which is of practical significance for spin current control.

In transition metal dichalcogenides $MX_2$ (M = Mo, W; X = S, Se, Te), with a layered honeycomb lattice and two inequivalent valleys in the $k$-space electronic structure in the hexagonal Brillion zone (BZ), due to the large separation of valleys in $k$-space and the resulting suppression of intervalley scattering, the valley index can be used in analogy to the spin in spintronics, opening a new research direction called 'valleytronics'.[13,14,15] Such a valley polarization achieved via valley-selective circular dichroism has been demonstrated theoretically and experimentally in those $MX_2$ systems without inversion symmetry (in monolayer cases or under an electric field).[13, 14, 16, 17, 18, 19, 20, 21, 22] However, a spin/valley current in $MX_2$ compounds caused by such a valley polarization has never been observed, nor its electric-field control. In this Article, using an optical excitation with circularly polarized radiation of $WSe_2$ electric-double-layer transistors, we present the first observation of a spin-coupled valley photocurrent whose direction and magnitude depend on the degree of circular polarization of the incident light and can be continuously controlled by the external electric field. It was found that due to the modulation of the degree of broken inversion symmetry, the two valleys can possess different optical selection rules for the generation of valley polarized photocurrent.

Similar to other layered $MX_2$ crystals, 2H-WSe$_2$ is composed of two formula units by stacking the Se-W-Se sandwiched structure (a monolayer unit) nonsymmorphically along the *c*-axis with $D_{3h}$ symmetry of each WSe$_2$ monolayer (Fig. 1c). A mirror operation in the $D_{3h}$ symmetry and the lack of inversion symmetry in monolayer form play a crucial role to induce a novel valley Zeeman-type spin splitting in the band structure.[22, 23] In bilayers and the bulk case, two adjacent monolayers are rotated by $\pi$ with respect to each other, making the whole structure centrosymmetric. Therefore their electronic states (bulk band structure shown in Fig. 1d) remain spin degenerate because of the inversion symmetry combined with time-reversal symmetry. As an effective way to induce inversion asymmetry in these bilayers or bulk WSe$_2$, applying a perpendicular external electric field can be used to regulate the band spin splitting, as shown in WSe$_2$ band structures with and without perpendicular electric field $E_{ex}$ [Figs. S1a and S1b in the Supplementary Information (SI), respectively]. Importantly, with inversion symmetry breaking, carriers in opposite valleys (with respect to the $\Gamma$ point) in the BZ have opposite spin angular momenta since the system is time-reversal-symmetry protected. This indicates that valley-dependent phenomena should be also spin dependent, implying the possibility of generating a spin current in WSe$_2$.

We fabricated WSe$_2$ single crystal flakes into electric-double-layer transistors (EDLTs), which have the capability to generate a large interfacial electric field to control electronic phases of solids[24, 25, 26, 27, 28, 29, 30] and modulate the spin texture in 2DESs[22, 31, 32]. Figure 2a is a cross-section diagram of a WSe$_2$ EDLT gated with ionic gel.[33] This large local interface electric field applied perpendicularly to the 2D plane can effectively modify interfacial band bending and the degree of inversion asymmetry. Owing to the band bending caused by the chemical potential alignment between the gel and the WSe$_2$, there is an electron accumulation with low carrier density at the gel/WSe$_2$ interface even before the external gate voltage $V_G$ is applied.

CPGE measurements induced by circularly polarized light on WSe$_2$ were used to detect a non-uniform distribution of photo-excited carriers and the generated spin current in $WSe_2$, with the configuration shown in Fig. 2a. The wavelength (1064 nm) used is smaller than the indirect band gap of $WSe_2$; thus the photocurrent generated originates from the surface accumulation layer, and not from bulk electron-hole excitations. To isolate the photocurrent response from a background current caused by laser heating gradients in the sample, we sweep the laser spot across two electrodes in the zero-biased $WSe_2$ EDLT device (Fig. 2b) with a fixed incident angle ($\theta = 60°$) and a fixed polarization for different heat gradients in the channel. The photocurrent switches its polarity as the laser spot is swept across the sample, but gives a non-zero finite value at the sample centre ($y = 0$) which would be a zero net current if the current only originates from the symmetric heating gradient. The observation of such a non-zero $j_y$ value at $y = 0$ gives us the indication that the generation of photocurrent might be caused by the non-uniform distribution of photo-excited carriers.

To confirm the dependence of the generated photocurrent on the helicity of the radiation light, the light polarization dependence of $j_y$ is measured at $y = 0$ (with different incident angles $\theta$, shown in Figs. 2d-i and Fig. S3). The first thing to be addressed here is the $\theta$–dependence of $j_{CPGE}$, showing the peak behavior (centered at around $\theta = 60°$ shown in Fig. 2c) is quite similar to the CPGE current observed in Rashba 2DESs. Secondly, one can see that when light is obliquely incident with a non-zero $\theta$, the obtained photocurrent $j_y$ exhibits a strong dependence on light circular polarization and oscillates with the rotation angle $\varphi$ of the quarter wave plate. This $j_y$, which can be quantitatively expressed as $j_y = C\sin 2\varphi + L\sin 4\varphi + A$, mainly includes two components, a π-periodic current oscillation term $j_{CPGE} = C\sin 2\varphi$ corresponding to the CPGE current, and a π/2-period oscillation term $L\sin 4\varphi$ corresponding to the linear photogalvanic effect (LPGE). The existence of $j_{CPGE} = C\sin 2\varphi$ component (red curves), satisfying the amplitude expression $j_{CPGE} = \eta\gamma I \sin\theta \sin 2\varphi$, directly reflects the helicity of the generated photocurrent. As indicated in Figs. 2d-2i, the direction and magnitude of this $j_{CPGE}$ strongly depend on the degree of circular polarization of the incident light, and $j_{CPGE}$ reverses its direction upon changing the radiation helicity from left-handed to right-handed.[8, 31]

The CPGE phenomenon and the spin photocurrent in Rashba 2DESs are highly sensitive to subtle details of the electronic band structure, and even a small band splitting may result in measurable effects. Therefore, the modification of the degree of inversion asymmetry with an external perpendicular electric field provides a simply way to control the CPGE photocurrent in this work. Figures 3 and S5 show the light polarization dependent $j_y$, obtained in a WSe$_2$ EDLT at varied external bias $V_G$ from zero to 1.1 V. As a common point, the magnitude of the electric current $j_y$ for all bias is related to the radiation helicity, and $j_{CPGE}$ (red curves) follows $C\sin 2\varphi$. Importantly, $j_{CPGE}$ dramatically increases with $V_G$ from tens of pA to thousands of pA (Figs. S3m and S3n), unambiguously indicating an electric modulation of the CPGE photocurrent. Such an electric modulation was very reproducible and observed in multiple devices.

To further understand the imbalanced electron distribution in $\boldsymbol{k}$-space and how $V_G$ modulates the spin-mediated photocurrent via tuning the degree of broken inversion symmetry, we compare the magnitude of photocurrent at $V_G = 0.3$ V under three special incidence angles, as shown in Figs. 4a, 4c and 4e. In most cases shown above and in Fig. 2a, where the photocurrent is generated transverse to the light scattering plane (*x-z* plane), the opposing angular-momentum polarizations that are excited by the different helicities must have an angular-momentum component in the *x-z* plane and be asymmetrically distributed along the *y*-direction in $\boldsymbol{k}$-space. When light is obliquely incident in the *y-z* plane, where azimuth angle $\phi = 90°$ (the angle between the *x* axis and the projection of incident light in the *x-y* plane, shown in Fig. 4c), one can see that the photocurrent completely disappears (Fig. 4d) because the device metal contacts lie in the light scattering plane. This indicates that the electrons involved in generating photocurrent have an angular momentum polarization which is locked perpendicular to their momentum. In the scenario with normally incident light (Figs. 4e and 4f), the photocurrent is negligibly small, as it is forbidden by the rotational symmetry ($C_{3v}$) about the normal axis. By considering these observations, these results reveal that the important features of helicity-dependent photocurrent $j_{CPGE}$ arises from the asymmetric optical excitation of the split bands.

We now theoretically show how the CPGE phenomenon is related to the valleys in the WSe$_2$ band structure. Since the Fermi energy can be tuned into the lowest conduction band with gel gating, the Fermi surface is located at six valleys around the $k_z = 0$ plane, as shown in Fig. 5a. The center of each valley (denoted as $\Lambda_i$ or $\Lambda_i'$ for convenience) lies in the $\Gamma K_i$ or $\Gamma K_i'$ direction (Fig. 5b). In atoms, optical transition selection rules are determined by the orbital angular momentum of the atomic levels. Bloch bands in solids can inherit these rules from their parent atomic orbits. With broken inversion symmetry, different valleys can be distinguished by the Berry phase of the Bloch bands,[14] which gives rise to the valley-dependent optical selection rule in WSe$_2$.

The photocurrent $j_{CPGE}$ can be derived by considering the light absorption of electrons occurring at six valley points $\Lambda_i$ and $\Lambda_i'$ in the lower conduction bands. We first consider the electron transition amplitude due to photon absorption at valley point $\Lambda_2$. The electron-photon interaction takes the form: $H' = \int dr^3 \frac{ie\hbar}{mc} \boldsymbol{A}(\boldsymbol{r}) \cdot \nabla$. To leading order, the momentum of the incident photon can be neglected compared to the electron momentum, and $\boldsymbol{A}(\boldsymbol{r}) = \boldsymbol{A}$ is therefore a constant vector. We denote the initial state in the lowest conduction band as $|I, \boldsymbol{k}_{\Lambda_2}\rangle$, and the final state in the higher conduction band as $|F, \boldsymbol{k}_{\Lambda_2}\rangle$. The transition amplitude is then of the form

$$\mathcal{M}_2(\boldsymbol{A}) = \mathcal{M}_{2,x} A_x + \mathcal{M}_{2,y} A_y + \mathcal{M}_{2,z} A_z = \boldsymbol{\mathcal{M}}_2 \cdot \boldsymbol{A},$$

where $\mathcal{M}_{2,u} = (e\hbar/mc)\langle F, \boldsymbol{k}_{\Lambda_2}|i\partial_u|I, \boldsymbol{k}_{\Lambda_2}\rangle$ for $u = x, y, z$. The current produced at $\Lambda_2$ is then

$$\boldsymbol{j}_{\Lambda_2} = \frac{\xi I}{A^2} \boldsymbol{e}_2 |\boldsymbol{\mathcal{M}}_2 \cdot \boldsymbol{A}(\theta, \phi)|^2,$$

where $\boldsymbol{e}_i = \boldsymbol{k}_{\Lambda_i}/|\boldsymbol{k}_{\Lambda_i}|$ is the unit vector along the $\Gamma K_i$ direction, $\theta$ and $\phi$ are the incident angle and azimuth angle of the light. $I$ is the incident light intensity, while $\xi$ is a coefficient related to the carrier density and velocities of the electron states. The currents produced at other valleys $\Lambda_i$ can be obtained through a $C_3$ rotation. The amplitude at valley $\Lambda_i'$ is related to that at $\Lambda_i$ through the time reversal symmetry $T$, $\boldsymbol{\mathcal{M}}_2' = -\boldsymbol{\mathcal{M}}_2^*$. The total current can therefore be expressed as:

$$\boldsymbol{j} = \frac{\xi I}{A^2} \sum_{i=1}^{3} \boldsymbol{e}_i \left( \left|\boldsymbol{\mathcal{M}}_2 \cdot \boldsymbol{A}\left(\theta, \phi - \frac{2(i-2)\pi}{3}\right)\right|^2 - \left|\boldsymbol{\mathcal{M}}_2^* \cdot \boldsymbol{A}\left(\theta, \phi - \frac{2(i-2)\pi}{3}\right)\right|^2 \right).$$

The condition for a non-zero $\boldsymbol{j}$ is $\boldsymbol{\mathcal{M}}_2^* \neq c\boldsymbol{\mathcal{M}}_2$, where $c$ is a phase factor (see SI). Our task then is to find $\boldsymbol{\mathcal{M}}_2$.

We note that the SOC, though strong in WSe$_2$, is not essential to generate the CPGE in a material with $C_{3v}$ symmetry. To show this, we set the SOC to zero for the moment, and see how the photocurrent arises just based on valley-dependent optical selection rules. The electron states near the Fermi surface are mainly composed of $d$-orbitals (Fig. S2). In the absence of the out-of-plane external electric field, the crystal has time reversal symmetry $T$, mirror reflection symmetries $R_x$ ($x \to -x$) and $R_z$ ($z \to -z$), and a non-symmorphic symmetry $Y = t_z^{1/2} \otimes R_y$ ($y \to -y$) that interchanges the two monolayers in a unit cell, where $t_z$ is the unit translation along the $z$-direction (Fig. 5c). This indicates the electron states at $\Lambda_2$ are eigenstates of the symmetries $R_z$, $Y$, and $TR_x$. In particular, $Y^2 = t_z = 1$ for $k_z = 0$, therefore $Y = \pm 1$. Combined with *ab initio* calculations (see Fig. S2 in SI), we find the initial state of the form $|I, \boldsymbol{k}_{\Lambda_2}\rangle = ia_0|d_{xy}^+\rangle + b_0|d_{x^2-y^2}^+\rangle + c_0|d_{3z^2-r^2}^+\rangle$, where $|d_{f(x,y,z)}^\pm\rangle$ denote the Bloch state $\sum_j e^{i\boldsymbol{k}_{\Lambda_2}\cdot \boldsymbol{r}_j}|d_{j,f(x,y,z)}^\pm\rangle$ with $Y = \pm 1$. Within reach of the photon energy, there are three states in the higher conduction bands (Fig. 5b), $|F_1, \boldsymbol{k}_{\Lambda_2}\rangle = ia_1|d_{xy}^+\rangle + b_1|d_{x^2-y^2}^+\rangle + c_1|d_{3z^2-r^2}^+\rangle$, $|F_2, \boldsymbol{k}_{\Lambda_2}\rangle = |d_{yz}^+\rangle$, and $|F_3, \boldsymbol{k}_{\Lambda_2}\rangle = ia_3|d_{xy}^-\rangle + b_3|d_{x^2-y^2}^-\rangle + c_3|d_{3z^2-r^2}^-\rangle$. All the coefficients $a_i$, $b_i$, $c_i$ are real, as is required by $TR_x$ symmetry. One can show that due to the $R_z$ and $Y$ symmetries, the only non-vanishing amplitude from $|I, \boldsymbol{k}_{\Lambda_2}\rangle$ to $|F_1, \boldsymbol{k}_{\Lambda_2}\rangle$ is the $x$ component $\mathcal{M}_{2,x}^{(I \to F_1)}$. This means $\mathcal{M}_2^{(I \to F_1)} = e^{2i\delta}\mathcal{M}_2^{(I \to F_1)*}$ where $\delta$ is the complex argument of $\mathcal{M}_{2,x}^{(I \to F_1)}$, and the transition from band $I$ to $F_1$ gives a net current zero. Similarly, the only non-vanishing amplitudes from $|I, \boldsymbol{k}_{\Lambda_2}\rangle$ to $|F_2, \boldsymbol{k}_{\Lambda_2}\rangle$ and to $|F_3, \boldsymbol{k}_{\Lambda_2}\rangle$ are $\mathcal{M}_{2,z}^{(I \to F_2)}$ and $\mathcal{M}_{2,y}^{(I \to F_3)}$ respectively, and therefore they also produce no net current (see SI for details).

So the net photocurrent is zero in the absence of the external out-of-plane electric field $E_{ex}$, which breaks inversion symmetry. Next we discuss how $E_{ex}$ generates the photocurrent and modulate its magnitude. When an out-of-plane electric field $\boldsymbol{E}_{ex} = E_{ex}\boldsymbol{e}_z$ is applied, the $WSe_2$ crystal loses the mirror reflection symmetry $R_z$. Since the two states $|F_1, \boldsymbol{k}_{\Lambda_2}\rangle$ and $|F_2, \boldsymbol{k}_{\Lambda_2}\rangle$ have opposite $R_z$ eigenvalues and are close in energy, this symmetry breaking induces a mix between the two states. For weak field $E_{ex}$, the two new eigenstates take the form $|F_1', \mathbf{k}_{\Lambda_2}\rangle = |F_1, \mathbf{k}_{\Lambda_2}\rangle + gE_{ex}|F_2, \mathbf{k}_{\Lambda_2}\rangle$ and $|F_2', \mathbf{k}_{\Lambda_2}\rangle = |F_2, \mathbf{k}_{\Lambda_2}\rangle - gE_{ex}|F_1, \mathbf{k}_{\Lambda_2}\rangle$, where $g$ is real as required by the $TR_x$ symmetry. Since in the above we have found the transition amplitude to $|F_1, \mathbf{k}_{\Lambda_2}\rangle$ and $|F_2, \mathbf{k}_{\Lambda_2}\rangle$, it is straightforward to show the transition amplitude to $|F_1', \mathbf{k}_{\Lambda_2}\rangle$ has two non-vanishing components $\mathcal{M}_{2,x}^{(I \to F_1')}$ and $\mathcal{M}_{2,z}^{(I \to F_1')} = m_{2,z}^{(I \to F_1')} E_{ex}$ (see SI). Due to the $TR_x$ symmetry, $\mathcal{M}_{2,x}^{(I \to F_1')}$ is restricted to be real while $\mathcal{M}_{2,z}^{(I \to F_1')}$ is purely imaginary. Therefore $\mathcal{M}_2^{(I \to F_1')*} \neq c\mathcal{M}_2^{(I \to F_1')}$, satisfying the condition to get a non-zero $\mathbf{j}$ in WSe$_2$ as we have observed in the experiments. The same analysis applies to the transition amplitude $\mathcal{M}_2^{(I \to F_2')}$ to $|F_2', \mathbf{k}_{\Lambda_2}\rangle$. These are the two channels inducing the CPGE. For circular polarized light obtained via a rotation angle $\varphi$ quarter wave plate with incident angle $\theta$ and azimuth angle $\phi$, the vector potential is $\mathbf{A}(\theta, \phi) = A[(\mathbf{e}_x \sin\phi - \mathbf{e}_y \cos\phi)(1 + i\cos 2\varphi) + i(\mathbf{e}_z \sin\theta - \mathbf{e}_x \cos\theta \cos\phi - \mathbf{e}_y \cos\theta \sin\phi)\sin 2\varphi]$. By a direct calculation one can show the net current $\mathbf{j}$ is given by

$$\mathbf{j} = \chi I E_{ex} \sin\theta \sin 2\varphi \, \mathbf{e}_\perp ,$$

where $\mathbf{e}_\perp = (\mathbf{e}_x \sin\phi - \mathbf{e}_y \cos\phi)$ is the in-plane unit vector perpendicular to the incident light, and the coefficient $\chi = 12g(\xi_1' - \xi_2')\left|\mathcal{M}_{2,x}^{(I \to F_1)} \mathcal{M}_{2,z}^{(I \to F_2)}\right|$ (see SI). Since the coefficient $\xi_i'$ is related to the electron filling controlled also by the gating voltage, the coefficient $\chi$ is also implicitly an increasing function of $E_{ex}$. The photocurrent can therefore be prominently tuned through the gating electric field $E_{ex}$.

We note that the current, which is perpendicular to the incident direction of the light, should be defined as a pure valley-polarized current without the need to consider the SOC.[9, 10, 34, 35] Once SOC is introduced, the spin degeneracy will be removed, and the transition amplitude of electrons with opposite spins will become different. The photocurrent $\mathbf{j}$ then becomes a partially spin-polarized photocurrent based on the valley polarized current (discussed in detail in SI), which gives us the intrinsic nature of the observed spin-coupled valley photocurrent.

Such generation and direct control of the spin-coupled valley photocurrent in a transistor configuration exhibit great potential toward the realization of practical applications, as well as physics for manipulating the spin and valley degrees of freedom in WSe$_2$. Given that this mechanism for achieving and modulating spin photocurrent should not be limited to these MX$_2$ dichalcogenides (like MoS$_2$, MoSe$_2$, or WS$_2$), our finding can be generalized to other 2DESs with valleys near the corners of a hexagonal BZ, thereby paving a new path towards developing spintronics/valleytronic devices.

## Methods

Two Ti/Au electrodes were fabricated on a freshly-cleaved WSe$_2$ single crystal flake (on SiO$_2$/Si wafer) with a channel of a few millimeters in length. Serving as the side gate electrode, a large-area Au pad was deposited near the WSe$_2$ but electrically insulated from the WSe$_2$. A typical EDL transistor was fabricated by drop-casting DEME-TFSI-based ionic gel [DEME-TFSI, N,N-diethyl-N-(2-methoxyethyl)-N-methylammonium bis-trifluoromethylsulfonyl)-imide from Kanto Chemical Co]. WSe$_2$ transistors can be obtained based on gel gating. The CPGE measurement configurations are shown in Fig. 2a and S3. The solid state laser with a wavelength 1064 nm (1.16 eV, smaller that the indirect band gap of WSe$_2$) was intentionally chosen in this study, to avoid the excitation from the valence band to the conduction band. To produce photocurrent, the laser beam with a 500 μm-diameter spot is incident on the device at the center between the two electrodes with the out-of-plane angle $\theta$ (defined as the angle of incident light from the normal direction of the $x$-$y$ plane). The helicity of laser light was modulated by a rotatable quarter wave plate $P_{circ} = \sin 2\varphi$. The polarization-dependent photocurrent was identified by measuring $j_y$ collected between two electrodes by AC lock-in amplifier. All the measurements on the spin photocurrent generation were performed at room temperature.

Electronic structure calculations were carried out by the Vienna Ab-initio Simulation Package

(VASP)[36, 37] within the framework of the Perdew-Burke-Ernzerhof-type generalized gradient approximation[38] of density functional theory.[39] Spin-orbit couplings are fully considered. The kinetic energy cut-off is set to 450 eV in all calculations. A $12 \times 12 \times 6$ grid is used for the k-mesh of the bulk calculation. The lattice constant and internal atomic positions are taken from experiments.[40] The external electric field was modeled by adding a dipole potential along the surface normal, i.e., (001) direction.

**Acknowledgements**

This work was supported by the Department of Energy, Office of Basic Energy Sciences, Division of Materials Sciences and Engineering, under contract DE-AC02-76SF00515. B.L., H.J.Z., G.X., H.Y.H. and S.C.Z. also acknowledge FAME, one of six centers of STARnet, a Semiconductor Research Corporation program sponsored by MARCO and DARPA. X.Q.W. and B.S. acknowledges the National Basic Research Program of China (No. 2012CB619300 and 2013CB921900) and the NSFC of China (No. 61225019, 11023003 and 61376060).

**Author Contribution**

H.T.Y., X.Q.W. and B.L. equally contributed to this work. H.T.Y., H.Y.H. and Y.C. conceived and designed the experiments. H.T.Y. performed sample fabrication and all optical and transport measurements. H.T.Y. X.F.F., X.Q.W. and B.S. performed CPGE measurements. B.L., H.J.Z., G.X., Y.X. and S.C.Z. performed all DFT calculations and theoretical analyses. S.C.Z., H.Y.H. and Y.C. supervised the project. H.T.Y., B.L. and H.J.Z. wrote the manuscript, with input from all authors.

**Author Information** The authors declare no competing financial interests. Correspondence and requests for materials should be addressed to hyhwang@stanford.edu (H.Y.H.), yicui@stanford.edu (Y.C.).

**Figure 1 Schematic diagrams of the circular photogalvanic effect (CPGE) based on Rashba spin splitting, and the crystal/electronic structure of layered 2H-WSe$_2$**

**a**, A schematic band diagram for spin-orientation-induced CPGE for direct optical transition in a 2DES with Rashba spin splitting. σ$^+$-excitation induces inter-subband transitions (yellow line) in the conduction band, where the spin splitting together with optical selection rules creates the unbalanced occupation of the positive ($k_y^+$) and negative ($k_y^-$) states and further yields spin-polarized photocurrent. **b**, A schematic diagram for the CPGE measurement. The helicity of laser light was modulated by a rotatable quarter wave plate with a relationship $P_{circ} = \sin 2\varphi$. The spin photocurrents depend on the helicity of the radiation field: it reverses its direction upon changing the radiation helicity from left-handed to right-handed. **c**, Top and side views of the layered structure of 2H-WSe$_2$. Note that the $D_{3h}$ symmetry ($C_{3v}$ +$M$) of each Se-W-Se monolayer includes mirror operations ($M$) which play an important role for the generation of the CPGE current. **d,** Electronic band structure of bulk 2H-WSe$_2$, where the conduction band minimum (CBM) sits at a non-symmetric point (defined as the Λ point) along the Γ-K-direction. When electrons are doped into the WSe$_2$ system, Fermi pockets will appear around the Λ-points.

**Figure 2 Schematic diagram and incident angular dependent CPGE measurement of ambipolar WSe$_2$ electric-double-layer transistors**

**a,** Schematic structure of a typical WSe$_2$ EDLT with ionic gel gating. By applying a gate voltage $V_G$ to the lateral Au gate electrode, ions in the gel are driven to the WSe$_2$ surface, forming a perpendicular electric field at the EDL interface. Note that even without an external bias, a relatively low carrier density accumulation layer exists at the WSe$_2$ surface. **b,** An incident position dependent photocurrent by sweeping the laser spot across the two electrodes in the zero-biased WSe$_2$ EDLT device with a fixed polarization. **c,** CPGE photocurrent $j_{CPGE}$ as a function of incident angle, showing a peak around $\theta = 60°$. **d-i,** Light polarization dependence of photocurrent $j_y$ in biased WSe$_2$ EDLT, measured at $y = 0$ with different incident angles $\theta$. Green circles are the measured $j_y$ following the form $j_y = C\sin 2\varphi + L\sin 4\varphi + A$. Blue dots are the photocurrent originating from the linear photo galvanic effect and obtained from the π/2-period oscillation term $L\sin 4\varphi$ by fitting. Red dots are the CPGE photocurrent with a π-periodic current oscillation.

**Figure 3 Electric field modulation of the spin photocurrent in WSe$_2$ EDLTs**

**a,** Electric field modulation of CPGE current $j_{CPGE}$ in WSe$_2$ EDLTs at various gate voltages $V_G$. **b,** CPGE current $j_{CPGE}$ as a function of $V_G$. The magnitude of the $j_{CPGE}$ can be modulated to a level above two orders larger than that of the zero bias case, providing a new way for a dramatic modulation of the spin photocurrent. The black, red and blue dots are from different samples.

**Figure 4 Light polarization dependent photocurrent in biased WSe$_2$ EDLTs with different incidence angles**

**a**, and **b**, Configuration and the CPGE measurement with the photocurrent generated transverse to the light scattering plane (*x–z* plane). Configuration and the CPGE measurement with light obliquely incident (at angle $\theta = 60°$) in the *y–z* plane shown in **c**, and **d**, and with normally incident light shown in **e**, and **f**. All measurement in this Figure are performed at $V_G = 0.3$ V.

**Figure 5 Physical origin of the generation of spin photocurrent in WSe$_2$ EDLTs and its electric field modulation**

**a**, The interpretation of the CPGE in reciprocal space. The hexagon represents the Brillion zone of WSe$_2$ in the $k_z = 0$ plane. The Fermi surface consists of six electron pockets (in blue) at $\Lambda_i$. When circular polarized light is applied, a photocurrent will be induced at each pocket $\Lambda_i$, yielding a net current *j* perpendicular to the incident light. **b**, Electronic band structure (conduction band) of 3D bulk WSe$_2$. CPGE arises when the Fermi energy is tuned into band *I*. The electrons are excited into higher bands $F_1$, $F_2$ and $F_3$. **c**, Illustration of the crystal symmetries of WSe$_2$. For simplicity only W atoms are plotted. The crystal has rotation symmetry $C_{3v}$, mirror reflection symmetries $R_x$ and $R_z$ with respect to the planes shown in the figure, and a non-symmorphic symmetry $Y = t_z^{1/2} \otimes R_y$ that interchanges the two layers.

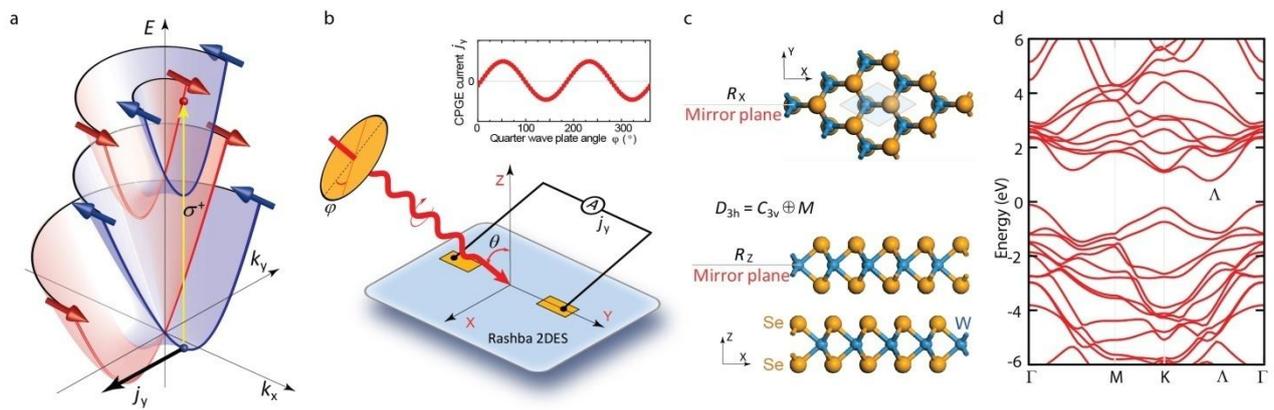

Figure 1

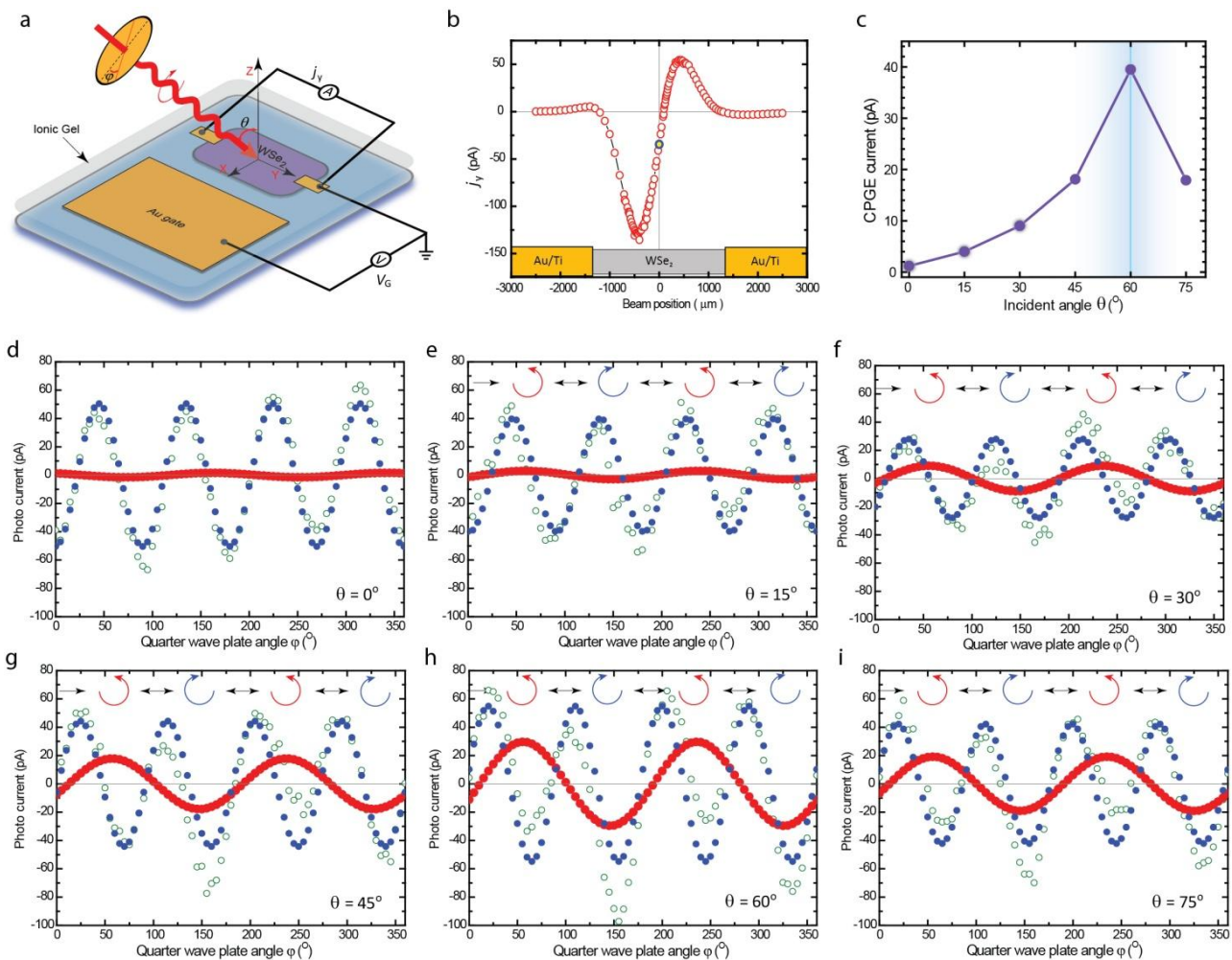

Figure 2

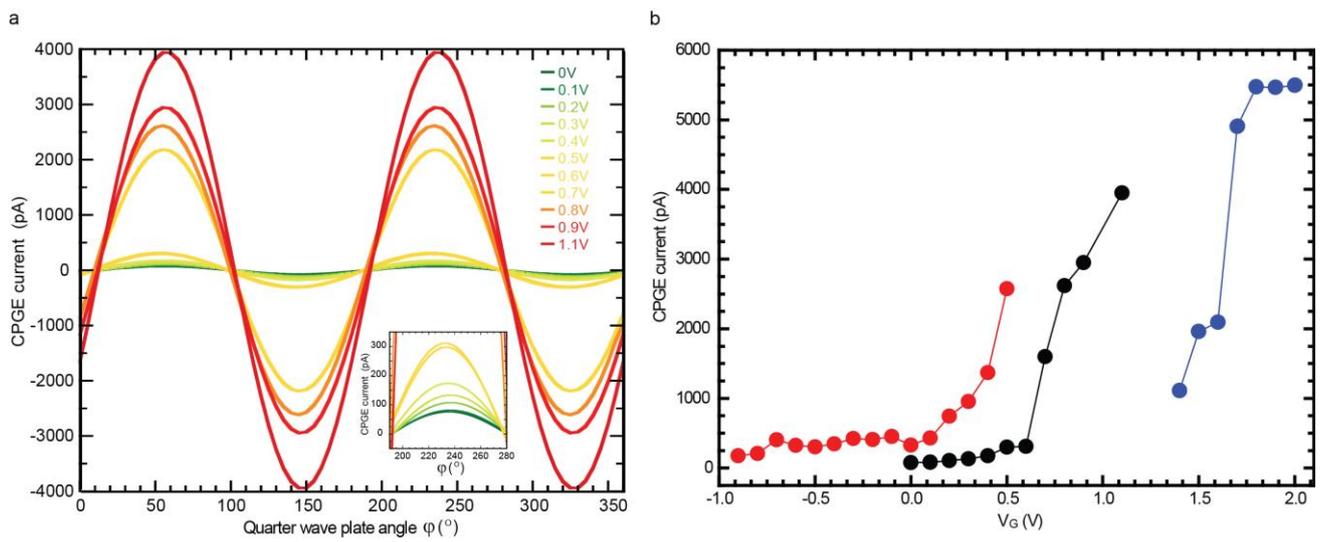

Figure 3

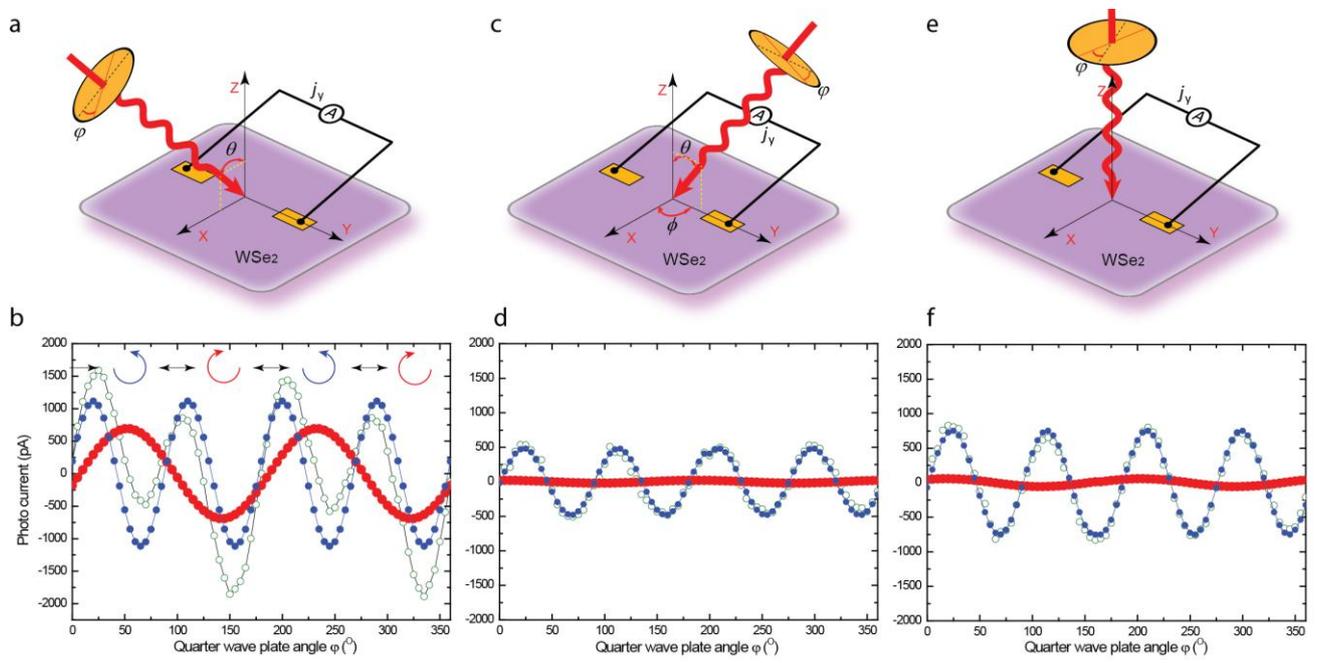

Figure 4

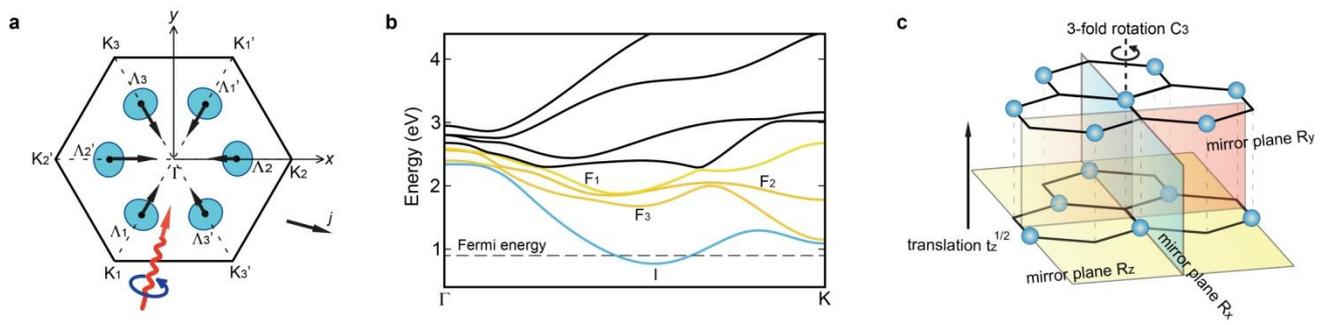

Figure 5

# Generation and Electric Control of Spin-Coupled Valley Current in WSe$_2$

Hongtao Yuan,[1,2] Xinqiang Wang,[3,4] Biao Lian,[1] Haijun Zhang,[1] Xianfa Fang,[3,4] Bo Shen,[3,4] Gang Xu,[1] Yong Xu,[1] Shou-Cheng Zhang,[1,2] Harold Y. Hwang,[1,2*] Yi Cui[1,2*]

[1]*Geballe Laboratory for Advanced Materials, Stanford University, Stanford, California 94305, USA*
[2]*Stanford Institute for Materials and Energy Sciences, SLAC National Accelerator Laboratory, Menlo Park, California 94025, USA*
[3]*State Key Laboratory of Artificial Microstructure and Mesoscopic Physics, School of Physics, Peking University, Beijing, 100871 China*
[4]*Collaborative Innovation Center of Quantum Matter, Beijing, China*

**Supplementary Information**

1. **Electronic band structure of WSe$_2$ thin film and its spin texture with/without an external electric field**

   To simulate the effect of an external electric field applied on the surface of WSe$_2$, *ab initio* calculations are carried out on a 3 unit-cell free-standing WSe$_2$ thin film with and without an electric field along the [001] direction. Figures S1a-c show the evolution of the band structures of the WSe$_2$ thin film before and after the application of an electric field. We can see that the conduction band minimum (CBM) is located around non-symmetric $\Lambda$-points along the $\Gamma-K$ direction, which implies that the Fermi surface should appear first at $\Lambda$-points and then $K$-points when the higher energy bands are filled with electrons under an electric field. The spin textures of the lowest two conduction bands are shown in Fig. S1d and S1e. The most important feature is that the spin texture is in-plane around the $\Gamma$-point, and yet out-of-plane in the other region in the BZ. So in this system, whether there is isotropic Zeeman spin splitting or the vortical Rashba spin texture strongly depends on whether the region is near $\Gamma$ or $K$ in the BZ. This spin texture indicates the existence of valley-spin coupling, and also that the valley-dependent phenomena are spin dependent.

---

[*] Corresponding author: hyhwang@stanford.edu, yicui@stanford.edu.

Some important characteristic features of the bands are stressed here: 1) The spin degeneracy is broken under an electric field. 2) Compared with the in-plane spin polarization around the zone center (high symmetric $\Gamma$-point) of the Brillouin zone (BZ), an out-of-plane Zeeman-type spin polarization takes place at low symmetric $\Lambda$-points (CBM sitting along $\Gamma$-$K$) and $K$-points (the corners of the BZ), in agreement with previous work[1]; 3) Rashba-type in-plane spin polarization at the $\Gamma$-point in conduction bands is extremely small and the lowest conduction band at the $\Gamma$-point is much higher than the CBM (~1.5 eV), indicating the obtained photocurrent does not involve the Rashba effect at the $\Gamma$-point; and 4) the energy splitting increases with increasing external electric field.

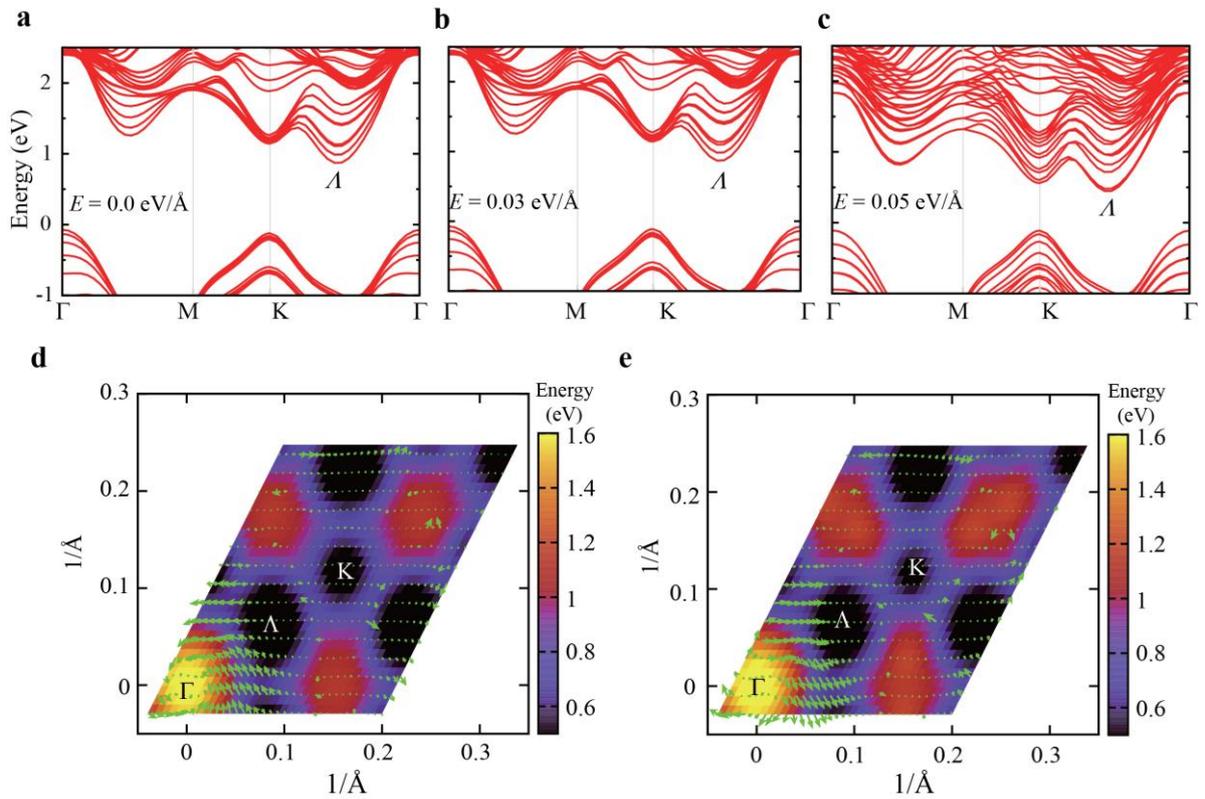

**Figure S1**. Evolution of band structure before and after applying an electric field. **a-c,** Band structure of a 3 unit-cell free-standing WSe$_2$ thin film with an external electric field $E_{ex}$ = 0 eV/Å in (a), 0.03 eV/Å in (b) and 0.05 eV/Å in (c). **d** and **e,** Spin textures of the 1$^{st}$ and 2$^{nd}$ –conduction subbands. Color bar indicates the energy level of each position in $k$-space. The green arrow represents for the direction and magnitude of the spin. They split from one spin-degenerate band, and have the opposite spin texture. The spin texture around the center of the BZ ($\Gamma$ point) is in-plane while it is out-of-plane near $\Lambda$ and $K$ points.

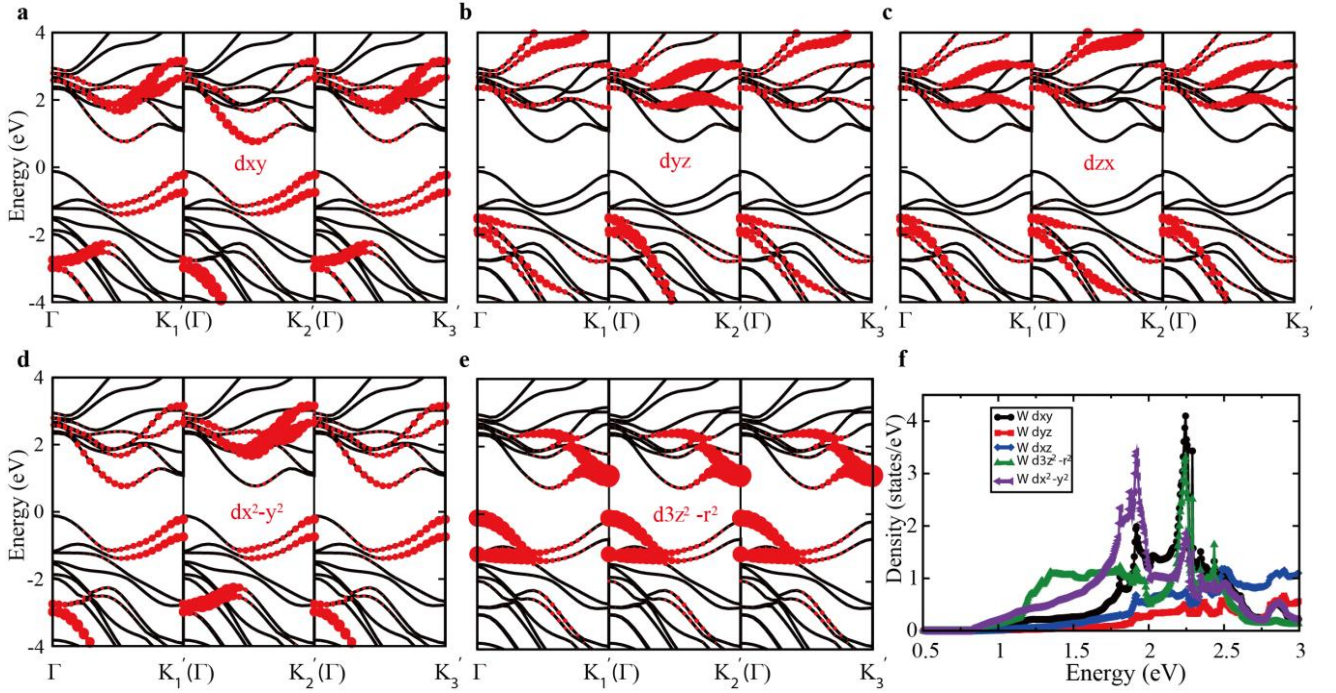

**Figure S2.** Atomic orbital character of the states along $\Gamma$ to $K$ (refer to Fig. 5a for notation). **a-e**, The band structure with the projection of all $d$ orbitals marked by the red dots. **f**, The projected density of states.

The projected band structure and density states of bulk WSe$_2$ are shown in Fig. S2. The bands near the band gap most come from $d$ orbitals of W. In particular, we note that the lowest conduction band consists of $d_{xy}$, $d_{x^2-y^2}$, $d_{3z^2-r^2}$ orbitals only. These projections onto different orbitals are useful in constructing our theoretical models discussed below.

2. **Optical setup for CPGE measurement**

As shown schematically in the experimental setup in Fig. S3, a quartz λ/4 plate is used to convert the incoming linearly polarized light (near infrared laser radiation with a wavelength of 1064 nm) into circularly polarized light. A chopper is used to modulated the light with a fixed modulating frequency of 233 Hz. Thus, the electrical signals with the same base frequency of the incident circularly polarized light can be obtained from the samples, which can be directly extracted by using standard AC lock-in techniques. The obtained photocurrent is expressed by $j_y = C\sin 2\varphi + L\sin 4\varphi + A$, where $j_{CPGE} = C\sin 2\varphi$ is the circular polarization contributed current, $j_{LPGE} = C\sin 4\varphi$ is the linear polarization contributed current, and $A$ is the polarization independent current. The same setup has been used for successfully measuring the circular photogalvanic effect for various Rashba systems based on group III-V semiconductors.[2] All measurements were performed in the regime of linear response, which was observed to extend at least up to ~0.5 W.

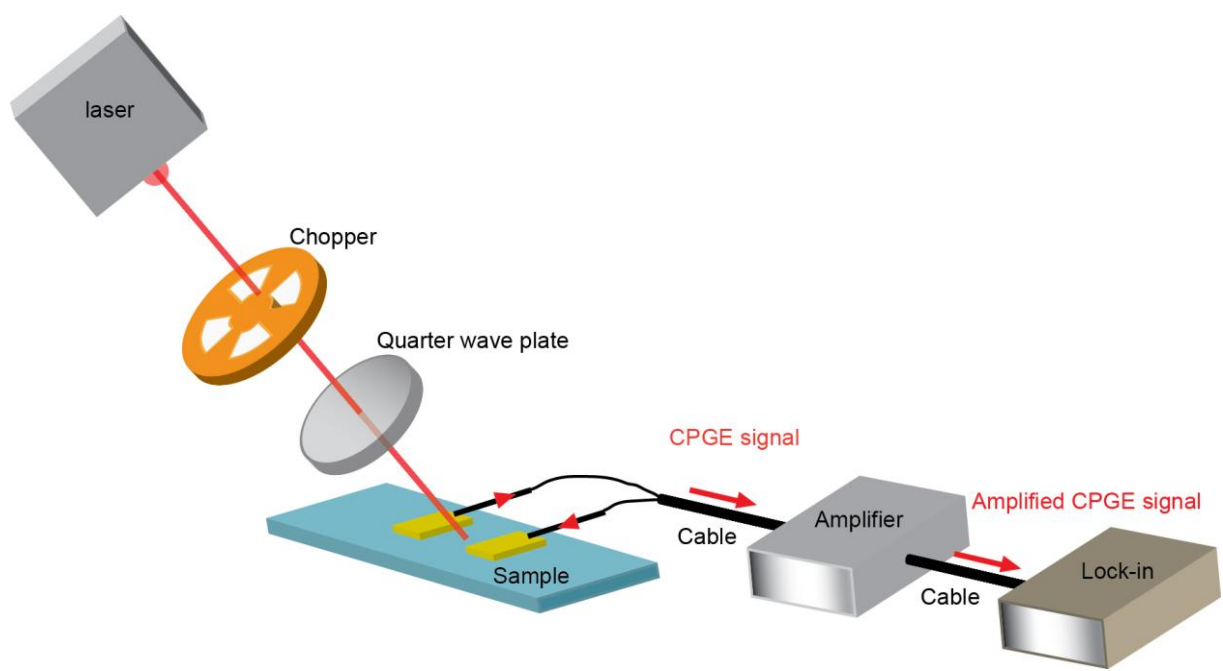

**Figure S3.** Schematic diagram of the CPGE measurement setup and the laser power dependence of the total photocurrent.

## 3. Transfer characteristics of WSe$_2$ EDLT for CPGE current

As shown in the transfer characteristics (sheet conductance $\sigma_{xx}$ as a function of gate voltage $V_G$) in Fig. S4, ambipolar transport of WSe$_2$ EDLTs with liquid gating can be achieved at room temperature.

One can clearly see a "U-shape" current response in the $I_{DS}$-$V_G$ plot, which suggests that electrons are accumulated at the interface when a positive $V_G$ is applied, while holes accumulate at a negative $V_G$. The relative large OFF current around zero-bias originates from the residual *n*-type (electron) conduction in the bulk crystal. Reducing the thickness of WSe$_2$ flakes can reduce the OFF current and further increase the ON-OFF ratio of the devices.[3] However no influence (no improvement) on the generation of the CPGE current is observed since the inversion symmetry breaking only occurs at the surface of the flakes, with no bulk contribution. To avoid any potential electrochemical reaction, we limit the bias within ±2 V, which is well within the chemical potential window (± 3 V for DEME-TFSI).

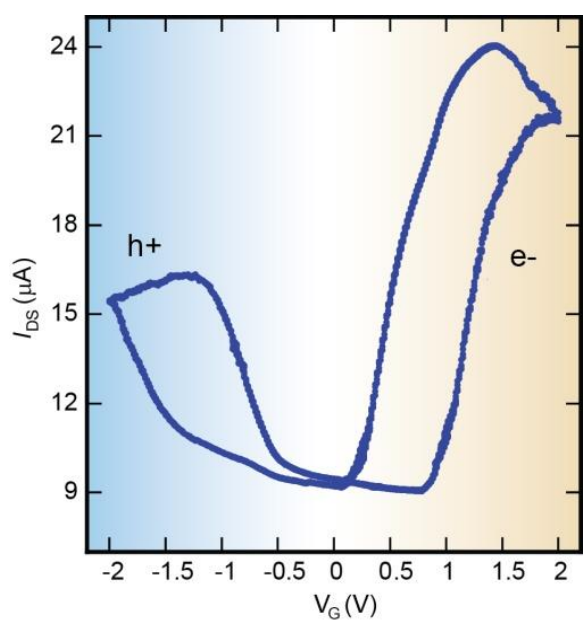

**Figure S4.** Transfer characteristic and ambipolarity of typical WSe$_2$ EDLTs.

The modification of SOC with an external perpendicular electric field provides a simply way to control the CPGE spin photocurrent. Figures S5a-S5l show the light polarization dependent $j_y$, obtained in a WSe$_2$ EDLT at varied external bias $V_G$ from zero to 1.1 V. As clearly seen, $j_{CPGE}$ dramatically increases with $V_G$ from tens of pA to thousands of pA (Fig. S3m and S3n), unambiguously indicating an electric modulation of the CPGE photocurrent.

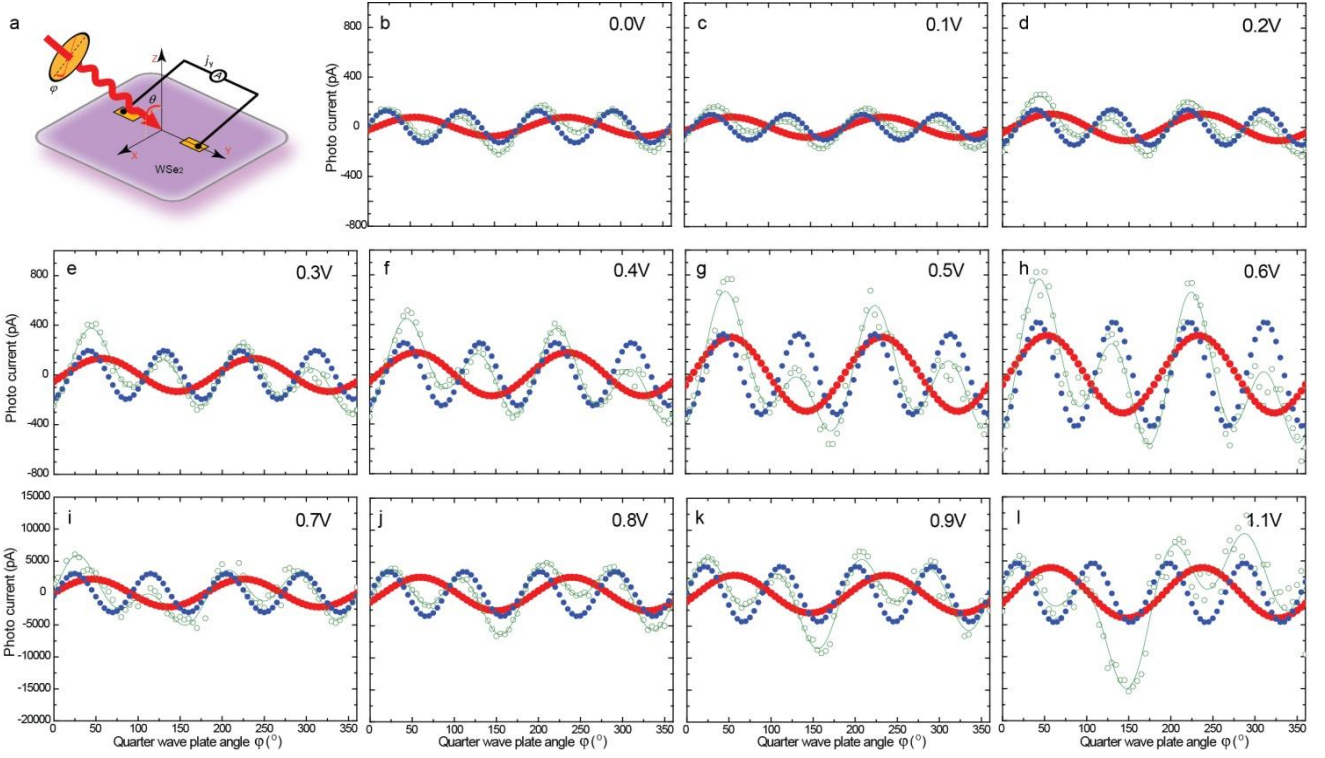

**Figure S5.** Bias dependent CPGE current in WSe$_2$ EDLTs with electron accumulation. Electric field modulation of photo current $j_y$ (green circles), CPGE current $j_{CPGE}$ (red dots) and LPGE current $j_{CPGE}$ (blue dots) in WSe$_2$ EDLTs at varied gate voltage $V_G$ from 0 V to 1.1 V. The solid thin green line is the fitting cure with the expression

$$j_y = C\sin 2\varphi + L\sin 4\varphi + A.$$

## 4. Valley optical selection rules and generation of pure valley current in WSe$_2$ without SOC

In the main text we show that the total current is

$$\boldsymbol{j} = \frac{\xi I}{A^2} \sum_{i=1}^{3} \boldsymbol{e}_i \left( \left| \boldsymbol{\mathcal{M}}_2 \cdot \boldsymbol{A}\left(\theta, \phi - \frac{2(i-2)\pi}{3}\right) \right|^2 - \left| \boldsymbol{\mathcal{M}}_2^* \cdot \boldsymbol{A}\left(\theta, \phi - \frac{2(i-2)\pi}{3}\right) \right|^2 \right),$$

where $\boldsymbol{\mathcal{M}}_2$ is the transition amplitude at $\Lambda_2$. We now prove the current is non-zero only if $\boldsymbol{\mathcal{M}}_2^* \neq c\boldsymbol{\mathcal{M}}_2$. We calculate explicitly:

$$|\boldsymbol{\mathcal{M}}_2 \cdot \boldsymbol{A}(\theta,\phi)|^2 - |\boldsymbol{\mathcal{M}}_2^* \cdot \boldsymbol{A}(\theta,\phi)|^2$$
$$= \left(\boldsymbol{\mathcal{M}}_2 \cdot \boldsymbol{A}(\theta,\phi)\right)\left(\boldsymbol{\mathcal{M}}_2^* \cdot \boldsymbol{A}^*(\theta,\phi)\right) - \left(\boldsymbol{\mathcal{M}}_2^* \cdot \boldsymbol{A}(\theta,\phi)\right)\left(\boldsymbol{\mathcal{M}}_2 \cdot \boldsymbol{A}^*(\theta,\phi)\right)$$
$$= \left(\boldsymbol{A}(\theta,\phi) \times \boldsymbol{A}^*(\theta,\phi)\right) \cdot \left(\boldsymbol{\mathcal{M}}_2 \times \boldsymbol{\mathcal{M}}_2^*\right).$$

Therefore, in order to have a non-vanishing current, one must have $\mathcal{M}_2^* \neq c\mathcal{M}_2$, and $\mathbf{A} \neq c'\mathbf{A}^*$.

The total current consists of the contributions of the transition from band $I$ onto several higher bands $F_i$, the amplitude of which is denoted by $\mathcal{M}_2^{(I \to F_i)}$. As is shown in the main text, through the first principle calculations and symmetry analysis, we find the initial state $|I, \mathbf{k}_{\Lambda_2}\rangle = ia_0|d_{xy}^+\rangle + b_0|d_{x^2-y^2}^+\rangle + c_0|d_{z^2}^+\rangle$, where $|d_{f(x,y,z)}^\pm\rangle$ is short for the Bloch state $\sum_j e^{i\mathbf{k}_{\Lambda_2} \cdot \mathbf{r}_j}|d_{j,f(x,y,z)}^\pm\rangle$ with $Y = \pm 1$. We now consider the transition to band $F_1$, which has a wave function $|F_1, \mathbf{k}_{\Lambda_2}\rangle = ia_1|d_{xy}^+\rangle + b_1|d_{x^2-y^2}^+\rangle + c_1|d_{z^2}^+\rangle$. Clearly, we have $Y|I, \mathbf{k}_{\Lambda_2}\rangle = |I, \mathbf{k}_{\Lambda_2}\rangle$ and $Y|F_1, \mathbf{k}_{\Lambda_2}\rangle = |F_1, \mathbf{k}_{\Lambda_2}\rangle$. This means that $\mathcal{M}_2^{(I \to F_1)}$ is invariant under symmetry operation $Y$. Also we have $R_z|I, \mathbf{k}_{\Lambda_2}\rangle = |I, \mathbf{k}_{\Lambda_2}\rangle$ and $R_z|F_1, \mathbf{k}_{\Lambda_2}\rangle = |F_1, \mathbf{k}_{\Lambda_2}\rangle$, so $\mathcal{M}_2^{(I \to F_1)}$ is invariant under $R_z$ as well. However, $\mathcal{M}_2^{(I \to F_1)}$ is a vector, therefore $\mathcal{M}_{2,y}^{(I \to F_1)}$ changes sign under $Y$, and $\mathcal{M}_{2,z}^{(I \to F_1)}$ changes sign under $R_z$. This means the only non-vanishing amplitude component is $\mathcal{M}_{2,x}^{(I \to F_1)}$. This means we have $\mathcal{M}_2^{(I \to F_1)} = \mathcal{M}_{2,x}^{(I \to F_1)} \mathbf{e}_x = e^{2i\delta}\mathcal{M}_{2,x}^{(I \to F_1)*}\mathbf{e}_x = e^{2i\delta}\mathcal{M}_2^{(I \to F_1)*}$. By the condition for producing current we derived above, this channel produces no net current. Following similar analysis, we can prove that the only non-vanishing amplitudes from $|I, \mathbf{k}_{\Lambda_2}\rangle$ to $|F_2, \mathbf{k}_{\Lambda_2}\rangle$ and to $|F_3, \mathbf{k}_{\Lambda_2}\rangle$ are $\mathcal{M}_{2,z}^{(I \to F_2)}$ and $\mathcal{M}_{2,y}^{(I \to F_3)}$ respectively. Similarly, all these channels have no contributions to the net current. The CPGE is therefore absent without external electric field.

When an out-of-plane electric field $E_{ex}$ is applied, as we have shown in the main text, band $F_1$ and $F_2$ mix to form two new states $|F_1', \mathbf{k}_{\Lambda_2}\rangle = |F_1, \mathbf{k}_{\Lambda_2}\rangle + gE_{ex}|F_2, \mathbf{k}_{\Lambda_2}\rangle$ and $|F_2', \mathbf{k}_{\Lambda_2}\rangle = |F_2, \mathbf{k}_{\Lambda_2}\rangle - gE_{ex}|F_1, \mathbf{k}_{\Lambda_2}\rangle$. For the transition from band $I$ to band $F_1'$, since band $F_1'$ satisfies $Y|F_1', \mathbf{k}_{\Lambda_2}\rangle = |F_1', \mathbf{k}_{\Lambda_2}\rangle$, we still have $\mathcal{M}_{2,y}^{(I \to F_1')} = 0$. However, we now have $\mathcal{M}_{2,x}^{(I \to F_1')} = \mathcal{M}_{2,x}^{(I \to F_1)} \neq 0$, and $\mathcal{M}_{2,z}^{(I \to F_1')} = gE\mathcal{M}_{2,z}^{(I \to F_2)} = m_{2,z}^{(I \to F_1')}E_{ex} \neq 0$. Further, under symmetry $TR_x$, we have $TR_x|I, \mathbf{k}_{\Lambda_2}\rangle = |I, \mathbf{k}_{\Lambda_2}\rangle$, $TR_x|F_1, \mathbf{k}_{\Lambda_2}\rangle = |F_1, \mathbf{k}_{\Lambda_2}\rangle$ and $TR_x|F_2, \mathbf{k}_{\Lambda_2}\rangle = -|F_2, \mathbf{k}_{\Lambda_2}\rangle$. By

applying a $TR_x$ symmetry operation, it is easily shown that $\mathcal{M}_{2,x}^{(I \to F_1')} = \mathcal{M}_{2,x}^{(I \to F_1')*}$ is real while $\mathcal{M}_{2,z}^{(I \to F_1')} = -\mathcal{M}_{2,z}^{(I \to F_1')*}$ is purely imaginary. We then have $\mathcal{M}_2 \times \mathcal{M}_2^* = 2\mathcal{M}_{2,x}^{(I \to F_1')}\mathcal{M}_{2,z}^{(I \to F_1')}\mathbf{e}_y \neq 0$. For the polarized light applied in our experiment, $\mathbf{A}(\theta,\phi) = A[(\mathbf{e}_x \sin\phi - \mathbf{e}_y \cos\phi)(1+i\cos 2\varphi) + i(\mathbf{e}_z \sin\theta - \mathbf{e}_x \cos\theta\cos\phi - \mathbf{e}_y \cos\theta\sin\phi)\sin 2\varphi]$, and it is straightforward to show $(\mathbf{A}(\theta,\phi) \times \mathbf{A}^*(\theta,\phi))_y = 2iA^2 \sin 2\varphi \sin\theta \sin\phi$, where circular polarization corresponds to angles $\varphi = (2n+1)\pi/4$. We then arrive at the final expression for the total current $\mathbf{j}^{(I \to F_1')}$ in this channel:

$$\mathbf{j}^{(I \to F_1')} = 8\xi_1' I \left(i\mathcal{M}_{2,x}^{(I \to F_1')}\mathcal{M}_{2,z}^{(I \to F_1')}\right) \sin 2\varphi \sin\theta \left(\mathbf{e}_1 \sin\left(\phi + \frac{2\pi}{3}\right) + \mathbf{e}_2 \sin\phi + \mathbf{e}_3 \sin\left(\phi - \frac{2\pi}{3}\right)\right)$$

$$= 12\xi_1' I E_{ex} \left|\mathcal{M}_{2,x}^{(I \to F_1')} m_{2,z}^{(I \to F_1')}\right| \sin\theta \sin 2\varphi \, \mathbf{e}_\perp$$

$$= 12g\xi_1' I E_{ex} \left|\mathcal{M}_{2,x}^{(I \to F_1)} \mathcal{M}_{2,z}^{(I \to F_2)}\right| \sin\theta \sin 2\varphi \, \mathbf{e}_\perp,$$

where $\mathbf{e}_\perp = (\mathbf{e}_x \sin\phi - \mathbf{e}_y \cos\phi)$ is the in-plane unit vector perpendicular to the incident light, and $\xi_1'$ is a coefficient for band $F_1'$. Adding contributions of transition from band $I \to F_2$ does not change the form of the net current. The final current can then be written as $\mathbf{j} = \chi I E \sin\theta \sin 2\varphi \, \mathbf{e}_\perp$, where the coefficient is given by $\chi = 12g(\xi_1' - \xi_2')\left|\mathcal{M}_{2,x}^{(I \to F_1)}\mathcal{M}_{2,z}^{(I \to F_2)}\right|$.

We emphasize that the newly arising non-vanishing amplitude $\mathcal{M}_{2,z}^{(I \to F_1')} = m_{2,z}^{(I \to F_1')} E_{ex}$ comes from the symmetry breaking of $R_z$ by the electric field, which is a purely orbital effect, and SOC is not required at all. We then expect an electric field controlled purely orbital galvanic effect to arise in materials with similar crystal structures but without significant SOC, like graphene, silicene, germanene or stanene.

5. **Valley optical selection rule considering the SOC and spin-coupled valley current in WSe$_2$**

The spin-orbital coupling term take the form $\lambda \mathbf{S} \cdot \mathbf{L}$, where the angular momentum $\mathbf{L}$ of d-orbitals is determined as

$$L_z = \begin{pmatrix} 0 & -2i & 0 & 0 & 0 \\ 2i & 0 & 0 & 0 & 0 \\ 0 & 0 & 0 & -i & 0 \\ 0 & 0 & i & 0 & 0 \\ 0 & 0 & 0 & 0 & 0 \end{pmatrix},$$

$$L_x = \begin{pmatrix} 0 & 0 & 0 & i & 0 \\ 0 & 0 & -i & 0 & 0 \\ 0 & i & 0 & 0 & 0 \\ -i & 0 & 0 & 0 & -\sqrt{3}i \\ 0 & 0 & 0 & \sqrt{3}i & 0 \end{pmatrix}, \quad L_y = \begin{pmatrix} 0 & 0 & i & 0 & 0 \\ 0 & 0 & 0 & i & 0 \\ -i & 0 & 0 & 0 & \sqrt{3}i \\ 0 & -i & 0 & 0 & 0 \\ 0 & 0 & -\sqrt{3}i & 0 & 0 \end{pmatrix},$$

under basis $(d_{x^2-y^2}, d_{xy}, d_{xz}, d_{yz}, d_{3z^2-r^2})$. Without the external electric field, the spin degeneracy is ensured by time reversal symmetry $T$ and the inversion symmetry $P = YR_xR_z$. This is because the combined operation $PT$ does not change the momentum but reverses the spin. However, due to the introduction of SOC, electron states may no longer be the eigenstates of the symmetry $Y$, and the $\pm$ index in states like $|d_{xy}^{\pm}\rangle$ are no longer exact. Through first principle calculations, we can obtain the orbital components and spin directions of each band (Fig. S1 and Fig. S2). For the lowest conduction band $I$, first principle calculations show two eigenstates at $\Lambda_2$ $|I, \boldsymbol{k}_{\Lambda_2}, \uparrow\rangle = |I, \boldsymbol{k}_{\Lambda_2}\rangle|\uparrow\rangle$ and $|I, \boldsymbol{k}_{\Lambda_2}, \downarrow\rangle = |I, \boldsymbol{k}_{\Lambda_2}\rangle|\downarrow\rangle$, where $|\uparrow\rangle$ and $|\downarrow\rangle$ are the up and down eigenstates of $S_z$, and they are related to each other through the $Y$ symmetry. Provided the orbital components of the wave functions of the bands are those as shown in the main text, their spin will be along the $z$ direction. This is because for band $I$ consisting of orbitals $d_{x^2-y^2}, d_{xy}, d_{3z^2-r^2}$, it is easily shown that the only non-vanishing orbital angular momentum is $L_z$, which is only coupled to $S_z$. The spin quantization axis is therefore the $z$ axis. Similar results apply to higher conduction bands.

When an out-of-plane external electric field $E_{ex}$ is applied, the inversion symmetry is lost, and the spin degeneracies will be lifted. As we have discussed in the main text, bands $F_1$ and $F_2$ will mix to form new states $|F_1', \boldsymbol{k}_{\Lambda_2}\rangle = |F_1, \boldsymbol{k}_{\Lambda_2}\rangle + gE_{ex}|F_2, \boldsymbol{k}_{\Lambda_2}\rangle$ and $|F_2', \boldsymbol{k}_{\Lambda_2}\rangle = |F_2, \boldsymbol{k}_{\Lambda_2}\rangle - gE_{ex}|F_1, \boldsymbol{k}_{\Lambda_2}\rangle$. Take state $|F_1', \boldsymbol{k}_{\Lambda_2}\rangle$ as an example, this induces an energy term $H_{F_1'} = \langle F_1', \boldsymbol{k}_{\Lambda_2}|\lambda \boldsymbol{S}\cdot \boldsymbol{L}|F_1', \boldsymbol{k}_{\Lambda_2}\rangle = 2\lambda a_1 gE_{ex}S_y$. The two spin states of band $F_1'$ then split in energy, and obtain an opposite spin $S_y$ component. We denote these two states as $|F_1', \boldsymbol{k}_{\Lambda_2}, +\rangle$ and $|F_1', \boldsymbol{k}_{\Lambda_2}, -\rangle$. (The spin degeneracy of band $I$ is also slightly lifted due to a mixing with band $F_2$, which we neglect here for simplicity.) The total current at $\Lambda_2$ includes the contributions of both $|F_1', \mathbf{k}_{\Lambda_2}, +\rangle$ and $|F_1', \mathbf{k}_{\Lambda_2}, -\rangle$, namely $\mathbf{j}_{\Lambda_2} = \mathbf{j}_{\Lambda_2,+} + \mathbf{j}_{\Lambda_2,-}$. Due to the spin splitting, $\mathbf{j}_{\Lambda_2,+}$ is no longer equal to $\mathbf{j}_{\Lambda_2,-}$. The spin current at $\Lambda_2$ can be defined as $\mathbf{j}_{\Lambda_2,S_y} = \mathbf{j}_{\Lambda_2,+} - \mathbf{j}_{\Lambda_2,-}$. The spin direction of the current is seen to be perpendicular to the current. Since the splitting of band $F_1'$ is proportional to $E$, we should have $|\mathbf{j}_{\Lambda_2,S_y}|/|\mathbf{j}_{\Lambda_2}| \propto E_{ex}$.

The above conclusion also applies for the total current and the total spin current, namely $|\mathbf{j}_S|/|\mathbf{j}| \propto E_{ex}$. The spin current $\mathbf{j}_S$ has the same direction as that of the total current $\mathbf{j}$, and has an in-plane spin polarization perpendicular to $\mathbf{j}_S$. Since $\mathbf{j}$ is proportional to the electric field $E_{ex}$, we find the spin current $\mathbf{j}_S$ proportional to $E_{ex}^2$.